\newcommand{\s}{\mbox{$\sigma $}}

\newcommand{\xd}{\mbox{$\dot{X}$}}
\newcommand{\xdm}{\mbox{$\dot{X^{\mu}}$}}
\newcommand{\zb}{\mbox{$\bar{z}$}}

\documentstyle[12pt]{article}

\newcommand{\ti}{\mbox {$    \tau _{1}$}}
\newcommand{\tw}{\mbox {$    \tau _{2}$}}

\newcommand{\be}{\begin{equation}}
\newcommand{\br}{\begin{eqnarray}}
\newcommand{\ee}{\end{equation}}
\newcommand{\er}{\end{eqnarray}}

\newcommand{\dds}{\mbox {$\frac{\delta}{\delta \sigma}$}}

\newcommand{\p}{\mbox {$ \partial$}}

\newcommand{\pp}{\mbox {$ \partial ^{2}$}}

\begin{document}
\title{Proper Time Formalism and Gauge Invariance in Open String
Interactions}
\author{B. Sathiapalan\\ {\em
Physics Department}\\{\em Penn State University}\\{\em 120
Ridge View Drive}\\{\em Dunmore, PA 18512}}
\maketitle
\begin{abstract}
The issue of gauge invariances in the sigma model
formalism is discussed at the free and interacting level.
The problem of deriving gauge invariant interacting
equations can be addressed using the proper time formalism.
This formalism is discussed, both for point particles and
strings.  The covariant Klein Gordon equation arises in
a geometric way from the boundary terms.  This formalism is
similar to the background independent open string
formalism introduced by Witten.
\end{abstract}
\newpage
\section{Introduction}

  An understanding of the gauge and other symmetries in string theory
is of the utmost importance in understanding the physical significance
of strings.  This is lacking at the moment.  From a practical point
of view we would like the symmetries to be manifest in the computational
scheme also.  An approach that looks promising to us in this respect
is the loop variable approach \cite{BS1} which is a
generalization of the sigma
model renormalization group method\cite{CL,CC,AS,FT,DS,JP,BM}.
However the work in
\cite{BS1,BS2} deals with the free theory.  One needs to extend it
to include interactions.  There  are several issues that arise:
one is the question of modifying the gauge transformations.
The second is the question of massive modes and finally there is the
issue of going off shell.  There is a well defined answer to these
questions in string field theory \cite{W1}
but we would like to approach it in
the loop variable framework because of the computational simplicity.
The loop variable approach was developed as an extension of
the results of \cite{BS3,BS4}
 to gauge invariant interactions.
In \cite{BS3} it was shown that the equations of motion
of the tachyon
in string
theory can be written as a proper time equation by
analogy with point particles.  The connection with the renormalization
group follows from the fact that the proper time $\tau$ in string theory
is related to the coordinate $z$ of the sigma model by
$z= e^{\tau + i \sigma}$ and so $\frac{d}{d\tau}$ is a generator
of scale transformations.  It was also shown
in \cite{BS3} that if one keeps a
finite cutoff one finds that instead of obtaining the low energy
non polynomial effective equations of motion where the massive
modes are
integrated out, one gets an equation
in which the massive modes are present and
which, for an appropriate choice of the cutoff,
is quadratic in the fields.
  For the special case of a tachyon we showed in \cite{BS4} that the off
  shell
3-tachyon vertex of string field theory can be reproduced if we keep a
finite cutoff.  In the language of vertex operators a finite cutoff
is equivalent to a hole of finite radius on the world sheet.  If one
lets the radius go to zero one recovers the usual punctured world
sheet.  In this case the vertex operator has to be of dimension (1,1)
or equvalently the particle has to be on shell.  If we keep a finite
radius, on the other hand, the particle can be off shell.
In the language of the renormalization group if one is far
away from the fixed point and one has all the irrelevant operators
then, effectively, you have a cutoff in the theory.  When the
cutoff goes to zero one is pushed towards the neighbourhood of
a fixed point where only the
marginal and relevant operators are present.  Conversely
if one is to keep a finite radius (cutoff) then one should keep all
 the massive modes. All this analysis has been done for the tachyon.

 If one
keeps track of the reparametrizations of the boundary of this hole
in the world sheet, then one needs extra variables in the theory
and it turns out that
this enables one to write down gauge invariant (free)
equations for the massive modes \cite{BS1,BS2}.
In order to extend the results obtained for the tachyon to higher
mass states what
needs to be done is to generalize
this construction to the interacting case.  Fortunately, for the massless
vector one does not need all this machinery to maintain gauge
invariance.  In this paper we concentrate on the massless case and
for simplicity we stay close to the mass shell.  It will
turn out that the proper time formalism can be extended to describe
this situation in a straightforward way. We will do it both for the
point particle and the string.  The results of \cite{BS3,BS4}
suggest that it should be possible to extend this off the mass shell
also.  We will also discuss briefly the propagation of
a gauge (point) particle.

  This paper is organized as follows:  In Section 2 we describe briefly
three different schemes for deriving free gauge invariant
equations in the sigma model formalism.  In Section 3 we describe
the proper time formalism for a particle in a background vector field.
The mechanism
of gauge invariance in the interacting case can be understood from this
example.  In Section 4 we extend this to strings and discuss the
mechanism of gauge invariance there.  In Section 5 we give some
concluding remarks and point out the similarity with Witten's
formulation of the background independent open string
equation.
\newpage
\section{Gauge Invariance in the Sigma Model Formalism}
\setcounter{equation}{0}
Let us describe three different ways of deriving the equations of motion
for a massless vector field, i.e. Maxwell's equations, in the open
string.  They each involve imposing some requirements on the vertex
operator:

\be
\int dz V(x) \equiv A_{\mu}(x)\p _{z} X^{\mu} \equiv
\int dz \int dk A_{\mu} (k) e^{ikX} \p _{z} X^{\mu}
\ee
\underline{Method I}:
  We require that $\dds V(x) \mid _{\sigma =0} =0$ where the \s
-  dependence  arises due to ultraviolet divergences that we usually
remove by
normal ordering .  Thus:
\be
V(x) = : V_{N}(x, \s ):
\ee
To get the \s - dependence of the expression in (2.1) a simple
method is to consider the vertex operator
$e^{i(kX + A \partial X)}$ and write it as
\be
exp(i(kX + A \p X) + \frac{k^{2}}{2} <XX> + A.k <X \p X> )
\ee
\[
= exp(i(kX + A \p X) + k^{2} \sigma + A.k \p \sigma )
\]
We have used $<XX> = 2 \sigma $ and $ <X \p X> = \p \sigma $ .
Expanding the
exponent and keeping the term linear in $A$ we get
\be
A_{\mu}(k) e^{ikX} \p _{z} X^{\mu}
= A_{\mu }(k) :e^{ikX} \p _{z}X^{\mu}
:e^{k^{2} \sigma }
\ee
\[
-ik.A:e^{ikX}:\p _{z} \s e^{k^{2} \sigma }
\]
 Varying w.r.t. \s   gives
\be
(k^{2} A_{\mu} (k) - k_{\mu} k.A ) :e^{ikX}\p X^{\mu}: =0
\ee
which is nothing other than $\p _{\mu} F_{\mu \nu} = 0$ in momentum
space. Note that the crucial point (for gauge invariance) in this
derivation is the fact that \s depends on $z$.
This is already a generalization of the usual $\beta$ - function method
where we require $ \frac{dV}{dln a} = 0$ , where $a$ is a fixed
cutoff.  One way to think of this is that
the flat world sheet cutoff $a$ is being replaced by $ae^{\sigma}$.
To lowest order in \s   this is sufficient.  To get results accurate
to higher orders one can replace the cutoff by the geodesic distance,
as has been done for instance in \cite{BE}. There are other ways
of obtaining the higher order pieces also.  Another crucial feature
is that in deriving (2.5) one has to perform an integration by
parts.  This assumes that there are no surface terms.  This will
not be true when we include interactions.

\underline{Method II} : We impose
\be
L_{0} V = 0 =   L_{1} V
\ee
where $L_{n}$ are the Virasoro generators. [$L_{n} V=
0 $ trivially
for $n>1$].
Naively this imposes two requirements on the vertex operator:
\be
k^{2}A^{\mu}=0  \, \, and \, \, k.A=0
\ee
the so called 'physical state' conditions.  However note that we have the
freedom to add to $V$ vertex operators of the form
\be
B(k) k . \p X e ^{ikX}
= B(k) \p _{z} e^{ikX}  = L_{-1}(Be^{ikX})
\ee
i.e. a total derivative.
Thus (2.7) becomes
\be
k^{2}A^{\mu} + k^{\mu} k^{2}B =0
\ee
and
\[
k.A + k^{2}B =0
\]
In the first equation we can replace
$k^{2} B $ by $-k.A$ and obtain eqn(2.5): $k^{2}A^{\mu} - k^{\mu}k.A=0$.
The role played by the Liouville mode is taken over by the
auxiliary field $B$.

\underline{Method III}
We require $ \{ Q, cV \} =0$ where $Q$ is the BRST operator and $c$
is the ghost (fermionic) field.  Using
\be
Q= \oint dz c(z)[-1/2 \p X \p X + \p c b ]
\ee
and $V$ as before we get
\be
\{Q,cV\}= 1/2(A.k \pp c c(z) - i k^{2} A_{\mu} \p X^{\mu} \p c(z) c(z))
\ee
Setting the RHS of (2.10) to zero we would get the usual physical state
conditions (2.7).  However we can add to $cV$ another operator of the
same dimension and ghost number:
\be
W = B(k) \p _{z} c e^{ikX}
\ee
and
\be
\{Q,W\}= (B c \pp c + ik \p X B c \p c ) e^{ikX}
\ee
Thus we should actually require that
$\{ Q, cV+W \} =0$ and this gives two equations:
\be
A.k/2 - B =0
\ee
and
\[
k^{2}/2 A^{\mu} - k^{\mu} B =0
\]
which, combined together, give Maxwell's equation.  Note that this
method is very similar to method II in that we need an auxiliary field
$B$.

  Each of these methods can be generalized to the massive cases as
well.  Before we describe that let us describe the gauge transformations.
  In method III it is obvious:
\be
\delta (cV) = [Q, \Lambda ]
\ee
where $\Lambda$ has ghost number zero, since $\{ Q, [Q, \Lambda ] \} =0$
identically (in 26 dimensions).  That is we can add to the vertex
operator $cV$ the piece $[Q, \Lambda ]$ and it does not affect the
BRST invariance properties.

  Thus letting  $\Lambda = \Lambda _{0} e^{ikX}$ we get

\be
[Q, \Lambda ] = cik^{\mu} \Lambda \p X^{\mu} e^{ikX} +
 k^{2}/2 \p c e^{ikX}
\Lambda
\ee
which gives
\be
\delta A_{\mu} = k_{\mu} \Lambda \, \, , \delta B = (k^{2}/2) \Lambda
\ee

  This method is obviously the  sigma model version of Witten's string
field theory equation\cite{W1}:
\be
Q \Psi = 0
\ee
and has the gauge invariance :
\be
\delta \Psi = Q \Lambda
\ee
  The generalization to higher mass levels is immediate - it is just a
matter of writing down the relevant vertex operators.  Although we
will not need it in this paper we will, for future reference,
give very briefly the results
for the next mass level.  The general vertex operator is
\be
W= [ S^{\mu} c \pp X^{\mu} + S^{\mu \nu} c \p X^{\mu} \p X^{\nu}
+ D \pp c +
\ee
\[
+ B^{\mu} \p c \p X^{\mu} + E c \p c b ] e^{ikX}
\]
The equations are $\{ Q, W \} =0 $.
\be
-(k^{2}/2 +1) S^{\mu} + B^{\mu} + ik^{\mu} D =0
\ee
\[
-S^{\mu} + ik^{\mu} S^{\mu \nu} + ik^{\mu} D + B^{\mu} =0
\]
\[
ik.S/3 + S^{\mu}_{\mu} /6 + D + 2/3 E =0
\]
\[
(1+ k^{2}/2 )D + ik.B/2 -3/2 E =0
\]
\[
(k^{2}/2 +1)S^{\mu \nu} - ik^{\mu} B^{\nu} + 1/2 \delta ^{\mu \nu} E=0
\]
and the gauge transformations are $[Q, \Lambda ]  $  with
\be
\Lambda = [\Lambda ^{\mu} \p X^{\mu} + \Lambda cb]
\ee
which gives:
\be
\delta S^{\mu} = \Lambda ^{\mu} - ik^{\mu} \Lambda
\ee
\[
\delta D = -ik.\Lambda /2 -3/2 \Lambda
\]
\[
\delta E = -(k^{2}/2 +1) \Lambda
\]
\[
\delta S^{\mu \nu} = i/2(k^{(\mu} \Lambda ^{\nu )} ) +1/2 \delta
 ^{\mu \nu} \Lambda
\]
\[
\delta B^{\mu} = (k^{2} /2 \, \, +1) \Lambda ^{\mu}
\]
(2.19) is invariant under (2.21) only in 26 dimensions.

In metod II the gauge transformation evidently corresponds to the
freedom of adding a piece $L_{-1} B e^{ikX}$ to the vertex operator
$A_{\mu} \p X^{\mu}
 e^{ikX}$.  The point is that this ambiguity is already
allowed for by the addition of (2.8) and hence {\em a fortiori}
is an invariance of the equations of motion.
 The generalization to higher mass levels
would be to add
\be
L_{-n} \Psi _{n}
\ee
to the vertex operator $V$ and then impose
\be
L_{m} (V+ \Sigma _{n} L_{-n} \Psi _{n} ) =0
\ee
The equations obtained on eliminating the $\Psi _{n}$ are
guaranteed to have gauge invariance of the form $V \rightarrow
V + L_{-n} \Lambda _{n}$ \cite{BP}
This is the sigma
 model version of the Banks-Peskin string field
theory.  Of course ,as shown there, this naive generalization , while it
has all the gauge invariances, does not correspond to string theory.
One has to get rid of many redundant fields and gauge invariances
associated with those fields.  The end result is a fairly involved
expression for the equation of motion\cite{BP}.
Nevertheless one could, if one
so desired, transcribe these results to the sigma model framework.

Finally, in method I gauge invariance corresponds to the freedom
to add total derivatives of the form $\p _{z} \Lambda(X)$
to the action (2.1).
The generalization to massive modes is what is described in detail in
\cite{BS1,BS2}. It involves introducing an infinite number of new
variables $x_{n}$ and vertex operators are expressed as derivatives
in $x_{n}$ rather than $z$.  The freedom to add total derivatives
in $z$ is generalized to that of adding total derivatives in $x_{n}$.
This method is closest in spirit to the renormalization group since
in the end we still require $\dds V =0$.  The gauge transformations
in this method are fairly simple\cite{BS1,BS2}.  We will not describe
it here since we are not going to discuss the massive modes.

In this section we have described three approaches to understanding
the issue of gauge invariance in the sigma model language, at the
free level.  We now have to generalize this to
the interacting level.  The BRST method (III) has been generalized
in the string field theory language to the interacting level \cite{W1}
and in a
form more closely related to
sigma model and two dimensional field theory \cite{AA,AC,W2,WL,W3}.
We are looking
for an analogous generalization for the first method.  At the
free level there appear to be certain advantages to this method and
the hope is that this may be true at the interacting level also.
In this paper we will restrict ourselves to the massless vector
(and the tachyon) - so we will not need the extra variables
used in the loop variable generalization of the first method.
\newpage
\section{The Proper Time Formalism and Gauge Invariance for Point
Particles}
\setcounter{equation}{0}

   The proper time formalism for free particles is well known
   \cite{F,S,N,M,SM,T}
     In \cite{BS3}
we modified it to describe a self interacting scalar particle.
It was then
shown that one could write a very similar equation for strings and this
led
directly to a proof of the proportionality of the equations of motion
and the $\beta$ - function (for the tachyon). Describing gauge
theories in the first quantized formalism is a little harder.  A lot of
work
has been done in applying the BRST formalism to this end \cite{SB}.
  In this
section
we want to describe a point particle in a background gauge field using
the proper time formalism.  We will also discuss briefly
the propagation of a gauge particle itself (albeit a free one) which is a
little  trickier.

The proper time equation for a massless free relativistic particle
is
\be
\frac {\p \phi [X, \tau]}{\p \tau} = \Box \phi [X, \tau ] =0
\ee
The solution to the first part of the equation is
\be
\phi [X, \tau ] = \int dX_{i} \int _{X(0) =X_{i}} ^{X(T)=X_{f}}
{\cal D} X  e^{i/2\int ^{T}_{0} d \tau ( \frac{\partial X}
{\partial \tau })^{2}}
\phi [X,0]
\ee
  The kernel in equation (3.2) is the evolution operator in
  proper time.  Integrating over $T$ from $0$ to $\infty$ sets
$\frac{d \phi }{d \tau} =0$ in eqn.(3.1) and gives us the
Klein Gordon propagator.  We will use (3.2) and require $ \frac{d \phi}
{d\tau} =0$ as in \cite{BS3}.  We can, if we want, now modify
the action to include various backgrounds and then requiring
$\frac{d \phi}
{d \tau} =0$ should give the required generalization of (3.1) to the
interacting equation.  In \cite{BS3}
 this
was done for a self interacting scalar field.

Following \cite{BS3} we write
\be
\phi (k',\tau ) = \int dk < e^{ik'X(\tau )}e^{ikX(0)}> \phi (k,0)
\ee
However unlike \cite{BS3}
the expectation is calculated using the action
\be
\int ^{T}_{0} d \tau [ 1/2 (\frac {\p X }{\p  \tau })^{2} +
A_{\mu} \frac {\p X^{\mu}}{\p \tau }]
\ee
The free two point function is given by :
\be
<X^{\mu} ( \ti ) X^{\nu} ( \tw ) > = \delta ^{\mu \nu} \mid \ti - \tw
\mid
\ee
To lowest order we get using momentum conservation
\be
\phi (k, \tau ) = e^{ k^{2} \tau } \phi (k,0)
\ee
Requiring $\frac{d \phi}{d \tau} \mid _{\tau =0} = 0$
 gives $k^{2} \phi =0$
- the massless Klein Gordon equation.  To next order we have to calculate
\be
\int ^{T} _{0} d \ti < e^{ik' X(\tau )}\xd ( \ti ) e^{ipX( \ti )}
e^{ikX (0)} >
\ee
In (3.7) we have written $A_{\mu} (x) \frac{\partial X^{\mu}}
{\partial \tau} $
as $\int dp A_{\mu}(p) e^{ikX(\tau )} \xdm $.  The
range of integration is restricted from 0 to $T$.  We can
simplify the calculation by exponentiating $\xd (\tau)$ into
$e^{i(p.X(\tau_{1}) + p_{1}. \dot{X} (\tau _{1}))}$ and we will
remember in the end to keep the piece linear in $p_{1}$.
We get, for (3.7),
\be
\int d \ti exp ( k'.p( \tau - \ti ) -k'.p_{1} + p.k \ti +
p_{1}.k + k'.k \tau )
\ee
The linear piece in $p_{1}$ gives
\be
(p_{1}.k - p_{1}.k') \int ^{\tau }_{0} d \ti exp((k'.p +k'.k)\tau
-k'.p \ti  + k.p \ti )
\ee
which in turn gives (using $k+k' +p =0$)
\be
(p_{1}.k -p_{1}.k') e^{-k'^{2} \tau}[\frac{ e^{(k.p-k'.p) \tau } -1}
{p.(k-k')}]
\ee
Setting $k'^{2}=0$ and requiring $\frac{d}{d\tau} \mid _{\tau =0} =0$
gives the piece (replacing $p_{1}$ with $A_{\mu} (p)$)
\be
(A.k - A.k')\phi (k) =(2A(p).k + A(p).p)\phi (k)
\ee
To next order we have to calculate
\be
\int ^{\tau}_{0} d \ti \int ^{\tau _{1}}_{0} d \tw
<e^{ik'.X(\tau )} e^{i (p.X(\ti ) + p_{1} \dot{X} (\ti ))}
e^{i(qX(\tw ) + q_{1} \dot{X} ( \tw ) )} e^{ik' X(0) }>
\ee
In calculating this expression we need correlators like

$<\xd (\tau _{1}) \xd (\tau _{2})>$
and it is important to keep track of the absolute value prescription
in (3.5) (otherwise the correlator vanishes).  To lowest order
in momentum we have
\be
\lim _{\epsilon \rightarrow 0}
p_{1}.q_{1} \int _{0}^{\tau} d \ti \int _{0}^{\ti} d \tw
<[\frac{X(\ti + \epsilon ) - X( \ti - \epsilon )}{2\epsilon}]
[\frac{X(\tw + \epsilon ) - X( \tw - \epsilon )}{2 \epsilon}]>
\ee
As long as $\tau _{2} < \ti - 2 \epsilon $
the correlator is zero.  Otherwise it gives
\be
\int _{\ti - 2 \epsilon}^{\ti} d \tw (2(\ti - \tw ) - 4 \epsilon )
=-4 \epsilon ^{2}
\ee
Thus (3.13) gives $-p_{1}.q_{1} \tau $ and acting on it with
$\frac{d}{d \tau}$ gives $-p_{1}.q_{1} $ or $-A^{2}$.  Adding all three
contributions gives $(i \p - A ) ^{2}\phi $ the Klein Gordon
equation in a background electromagnetic field.
The other pieces from (3.12) give zero when we act with
$\frac{d}{d \tau } \mid _{\tau =0}$ on them.

  From (3.4) one can see that the construction is gauge invariant.
The transformation $A_{\mu} \rightarrow A_{\mu} + \p _{\mu} \Lambda$
does not leave the action invariant but
results in a boundary term :
\be
\int _{0} ^{T} d \tau \xd \frac{d \Lambda }{d X} = \Lambda (T) -
\Lambda (0)
\ee
This results in a phase,
which can be compensated by a gauge transformation
\be
\phi (\tau ) \rightarrow e^{i\Lambda (\tau )} \phi (\tau )
\ee
As explained in the last section, gauge invariance at the free level
is due
to the freedom to add total derivatives.  However if there are
boundary terms then the action is not invariant.  This is the
situation when one has interactions.
We then have to compensate by the transformation (3.16).  This is
the origin of inhomogeneous terms , i.e. those of the form
$\delta \phi = i \Lambda \phi $, (as against terms of the
form $\delta A_{\mu} = \p _{\mu} \Lambda $) - they arise from
boundaries of the integration region.
It is not obvious in the calculation of the covariant Klein Gordon
equation that the interaction terms  $ A_{\mu } \p ^{\mu} \phi ,\,
\p . A  \phi \, , \, A.A \phi $
also arise in this manner (from surface terms), but this is in
fact the case.
In the next section we will repeat
 the calculation in a way that makes this
fact manifest.

  One can now
  ask the following question: We understand how gauge invariance
is maintained as far as background gauge fields are concerned.  What
about deriving equations of motion
for the gauge particle itself (i.e. Maxwell's or Yang Mills equations)
in this formalism?
This is a little tricky since we do not usually treat the electromagnetic
field in first quantized form.  However motivated by strings we can
extend the previous discussion  and consider an object of the form
\be
<k_{1}.\xd (\tau ) e^{ik.X(\tau )} A_{1}.\xd (0) e^{ip.X(0)}>
\ee
and require $\frac{d}{d \tau} \mid _{\tau =0}=0$ as before.
\footnote{In string theory \xd acts on the ground state
and excites it to a vector state.  There is no such interpretation
for a point particle.  Perhaps we can think of $\xd \mid 0>$
as a current source for a photon.  For our purposes
we will not worry about interpreting it but
we will formally treat it just as in string theory since that is our
real interest in any case.}  We immediately run into a problem - that
of gauge invariance.  In eqn.(3.4) the vertex operator $\xdm (\tau)$
was integrated over  $\tau$.  So it was a gauge invariant expression
(except
for surface terms which we took care of by transformong $\phi$).
$\xdm (0) $  in the unintegrated form
has no such gauge invariance.  We will therefore modify
(3.17) to
\be
\int d \ti \int d \tw
<k_{1}.\xd (\tau ) e^{ik.X(\tau )} A_{1}.\xd (0) e^{ip.X(0)}>
\ee
This construction is gauge invariant but now the proper time equation
makes no sense - since \ti   and \tw   are both integrated over.
One must generalize the proper time prescription.  We can do as follows:
We know that $<X( \tau) X(0)>= \mid \tau \mid $.  Let us treat the
entity $<X(\tau ) X(0)>$ as a {\em field} $\Sigma ( \tau )$ and require
$\frac{\delta}{\delta \Sigma} =0$.  Here $\Sigma$ plays the same
role as the Liouville mode $\s$ in section 2.  As in sec.2 the integrals
$\int d \ti \int d \tw $  allow us to
integrate by parts.  In that case (3.18) gives
\be
\int d \ti \int d \tw [
k_{1}.A (p) \p _{\ti} \p _{\tw}
<X(\ti ) X(\tw )>
\ee
\[
+k_{1}.p A.k \p _{\ti}
<X(\ti ) X(\tw )>
\p _{\tw}
<X(\ti ) X(\tw )> ]
e^{k.p
<X(\ti ) X(\tw )> }
\]
\[
=
\int d \ti \int d \tw [
k_{1}.A (p) \p _{\ti} \p _{\tw}
\Sigma ( \ti - \tw )
\]
\[
+k_{1}.p A.k \p _{\ti}
\Sigma ( \ti - \tw )
\p _{\tw}
\Sigma ( \ti - \tw )]
e^{k.p
\Sigma ( \ti - \tw )}
\]
Varying w.r.t $\Sigma$ gives
\be
(k_{1}.A k.p - k_{1}.p A.k )\p _{\ti } \p _{\tw} \Sigma ( \ti - \tw )
e^{k.p \Sigma ( \ti - \tw )}
\ee
Set $p_{0} = - k_{0} $ (momentum conservation) and look at the
coefficient of $k_{1}^{\mu}$: It gives Maxwell's equation
$\p _{\mu} F^{\mu \nu} =0$.  The same method obviously
works for strings also
since we never needed the explicit form of the two point function of $X$.
\newpage

  To summarize this section, we have derived the gauge invariant
equation for a scalar using the proper time method.
  We have also shown how the proper time formalism can be
used for gauge particles at the free level.  Both these
can be immediately generalized to strings.
\newpage
\section{Proper Time Formalism and Gauge Invariance for Strings}
\setcounter{equation}{0}
We now apply the proper time formalism to strings: Replace $\tau$ by
$ln z$ to get
\be
[\frac{d}{dlnz} -2]<e^{ik'X(z)} e^{ikX(0)}>\phi (k) =0
\ee
In sec.2 we derived equations of motion by requiring that the
vertex operator have dimension one.
In eqn.4.1 we have two vertex operators and so it has dimension two
and hence should fall off as $1/z^{2}$ as equation (4.1) indicates.
We will calculate the expectation value using the action
\be
1/2 \int d^{2}z \p _{z}X \bar {\p} _{\zb}X + \int _{0}^{w}
 A_{\mu} \p _{z}
X^{\mu}
\ee
The action has the gauge invariance
\be
A_{\mu} \rightarrow A_{\mu} + \p _{\mu} \Lambda \, ,  \, \phi
\rightarrow e^{i\Lambda } \phi
\ee
as in the point particle case.
The two point function is :
\br
<X(z_{1}) X(z_{2})> & = & ln (z_{1}-z_{2}) , z_{1} \neq z_{2} \\
                    & = & ln(ae^{\sigma})  , z_{1}=z_{2}
\er
However we will just leave it as $<X(z_{1}) X(z_{2})>$ till the end of the
calculation.  To lowest order we get from (4.1) $(k^{2} -2) \phi$.
At the next order we have
\be
<e^{ik'X(z)} \int _{w} ^{z} dz_{1} A_{\mu} \p _{z} X^{\mu} (z_{1})
e^{ikX(z_{1})}
e^{ipX(w)}>
\ee
which gives
\be
\int _{w}^{z} dz_{1} [iA.k'\p_{z_{1}}
<X(z)X(z_{1})>
+iA.p\p_{z_{1}}
<X(z_{1}) X(w)>
]
\ee
\[
exp (k.k'
<X(z)X(z_{1})> +
k.p
<X(z_{1}) X(w)> +
k'.p
<X(z) X(w)> )
\]
To lowest order we get the surface terms:
\be
iA.k'[<X(z) X(z) - <X(z)X(w)>] +
\ee
\[
iA.p[<X(z)X(w)>-<X(w)X(w)>]
\]
\[
=-i(A.k' -A.p)ln(\frac{z-w}{a})
\]
This contributes $-i(A.k' - A.p)$ to the equation of motion.

At the next order we have
\be
<e^{ik'X(z)}\int _{w} ^{z} du \int _{w}^{u} dv A(k)\p X(u) e^{ikX(u)}
A(q)\p X(v) e^{iqX(v)}e^{ipX(w)} >
\ee
Again to lowest order in momenta we get
\be
\int _{w}^{z} du\int _{w}^{u} dvA(k)A(q)<\p _{u} X(u) \p _{v} X(v)>
\ee
\[
=\int _{w} ^{z} A(k)A(q) [<\p _{u} X(u) X(u)> - <\p _{u} X(u) X(w) >]
\]
\[
=\int _{w} ^{z} du A(k)A(q)
[1/2 \p _{u} <X(u) X(u)> - \p _{u} <X(u) X(w)>]
\]
\[
=A(k)A(q)[1/2[<X(z)X(z)>- <X(w)X(w)>]
\]
\[
-<X(z)X(w)> + <X(w)X(w)>]
\]
\be
=A.Aln(\frac{z-w}{a})
\ee
Adding up all the pieces we get $(\p - A ) ^{2} \phi = 0$
In following the steps from (4.6) to (4.10) one can see how
each contribution
is the surface term
 in an integral and how they conspire to reproduce
the gauge invariance as described in eqn.(4.3).  All this works
exactly the same way as for the point particle since we never really
needed to know the functional form of the two point function.  In
fact as indicated at the end of the last section we could have
just required $\frac{\delta}{\delta <X(z)X(w)>} = 2$ instead of
$\frac{d}{dln(z-w)} =2$.

In this section we have concentrated on understanding
the features that are
common to particles and strings, in particular, those that deal
with the massless gauge invariance.  We have shown that the
proper time formalism can be made gauge
 invariant.
\footnote{We can derive Maxwell's equation also in the string case
just as was done at the end of the last section by requiring
$\frac{\delta}{\delta <X(z)X(w)>} \int dz \int dw < \p _{z} X e^{ik.X}
\p _{w} X e^{ip.X}> = 0$.}
In this section we kept only the lowest order (in momentum) terms.
For point particles if we had similarly kept only the lowest
order terms the result (i.e. the Klein-Gordon equation)
would still be exact, as the calculation in Section 3 shows.
Thus the higher order terms must vanish. This is not so for
strings, however.  There are higher order corrections
to the Klein Gordon equation that ought to be evaluated.  Some of
these have
been calculated in various approximation schemes\cite{FT2,AA,AC}.
It should be possible, however, to do it in a systematic
way where the degree to which the massive modes are integrated out
can be controlled.  The parameter that controls this would be
the cutoff of the two dimensional field theory.  The
proper time formalism \cite{BS3,BS4}
appears to be a way of implementing this
idea.

\newpage
\section{Conclusion}
\setcounter{equation}{0}

In this paper we have attempted to understand gauge invariance
in the framework of the renormalization group both at the free level
and interacting case.  Our aim is to have an understanding at
the computational level rather than a formal proof of gauge
invariance. To this end we have made some progress in understanding
gauge invariance of the massless particle at the interacting level
provided we stay close to the mass shell.
One can also address these questions in the
BRST framework.  We saw in the second section  the similarities
between the two approaches at the free level.  In fact proceeding
to the interacting theory we can see that eqn.(4.1) is very similar
to the equation based on  the Batalin-Vilkovisky formalism used
in \cite{W2,WL,W3}.  Instead of $d/dlnz$ acting on the two point
function
one can have $Q
_{BRST}$ act on it. Witten's anti bracket is
essentially the Zamolodchikov metric-the two point function.  If we
were to include ghosts and use $cV$ instead of $V$ in (4.1) ($c$ being
the reparametrization ghost) we would have Witten's antibracket.  In
fact we have already seen in Sect3 that when dealing with gauge particles
the vertex operator should be integrated over.  Thus we should have
$\int dz V$ (which has the same dimension as $cV$).  Thus this formalism
seems very similar to that of \cite{W2,WL,W3}.

             We would like to extend the results of this paper
by going off shell and including the massive modes.
This issue can be hopefully addressed in this
formalism, just as was done for the case of the tachyon, by keeping
a finite cutoff.  As we change the value of the cutoff one should
be able to interpolate continuously from a string field
theory where all the modes are present to a low energy effective action
obtained via the sigma model formalism.  Presumably the extra coordinates
of \cite{BS1} will need to be introduced to maintain reparametrization
invariance.  We hope to return to these questions.

\underline{Acknowledgement}: I would like to thank W. Siegel
for many useful discussions. Most of this work was done while the
author was visiting the Institute for Theoretical Physics at
Stony Brook. I would like to thank the members of the ITP, and
especially M. Rocek,
for their hospitality.

\end{document}